\documentclass[11pt, a4paper]{scrartcl}
      
\usepackage{amsmath}
\usepackage{amsfonts}
\usepackage{amsthm}
\usepackage{amssymb}           
\usepackage{exscale}
\usepackage{cite}

\theoremstyle{definition}

\newcommand{\mC}{\mathbb{C}}
\newcommand{\mZ}{\mathbb{Z}}

\newcommand{\ket}[1]{|#1\rangle}

\newcommand{\ibar}{\bar{\imath}}
\newcommand{\jbar}{\bar{\jmath}}

\newcommand{\zetabar}{\bar{\zeta}}
\newcommand{\Iket}[1]{|#1\rangle\!\rangle}

\newcommand{\Bket}[1]{|\!|#1\rangle\!\rangle}

\newcommand{\mH}{\mathcal{H}}

\newcommand{\tpsi}{\tilde{\psi}}

\newcommand{\tlambda}{\tilde{\lambda}}

\newcommand{\sN}{\mathcal{N}}
\newcommand{\aob}[2]{\genfrac{}{}{0pt}{}{#1}{#2}}

\DeclareMathOperator{\mTr}{Tr}

\renewcommand{\paragraph}[1]{
  \vspace{0.5cm}
  \pagebreak[1]
  \noindent
  {\it #1}
  \nopagebreak
  \vspace{0.3cm}
  \nopagebreak}

\numberwithin{equation}{section}

\allowdisplaybreaks[1] 

\begin{document}

\begin{titlepage}
\renewcommand{\thefootnote}{\fnsymbol{footnote}}
  \flushright{hep-th/0504196}

  \begin{center}
    
    \vspace{3cm}
    {\LARGE \textbf{BPS branes in discrete torsion orbifolds}}
    
    \vspace{1cm} 
    {\Large Hanno Klemm\footnote{\tt{klemm@phys.ethz.ch}}}

\vspace{1cm}

    \emph{Institut f\"ur Theoretische Physik, ETH  H{\"o}nggerberg
      \\       8093 Z\"urich, Switzerland\\ and   \\
      Department of Mathematics, King's College London, \\
          Strand, London WC2R 2LS, UK }
    
    \vspace{5cm} 

    \begin{center}
      \textbf{ Abstract}
    \end{center}
\begin{abstract}
  We investigate D-branes in a $\mZ_3\times\mZ_3$ orbifold with
  discrete torsion. For this class of orbifolds the only known objects
  which couple to twisted RR potentials have been non-BPS branes.  By
  using more general gluing conditions we construct here a D-brane
  which is BPS and couples to RR potentials in the twisted and in the
  untwisted sectors.
\end{abstract}

\end{center}
\end{titlepage}

\section{Introduction}
\label{sec:introduction}

Orbifolds \cite{DHVW} have been considered extensively in string
theory as interesting backgrounds which capture a lot of the geometric
information of Calabi-Yau backgrounds while still being sufficiently
easy to be treated exactly by CFT methods. Shortly after the
introduction of orbifolds, Vafa introduced a slight generalisation of
the concept of an orbifold which is called a discrete torsion orbifold
\cite{Vafa:1986wx}. D-branes on orbifolds have been considered in a
plethora of contexts, and are by now quite well understood. For some
early works see for example \cite{Quivers,GDD}.  D-branes on orbifolds
with discrete torsion have been considered less frequently in the
literature and some puzzles still remain.  Some previous work on
D-branes in discrete torsion orbifolds can be found in
\cite{Douglas:1998xa,Douglas:1999hq,Mukhopadhyay:1999pu,Sharpe:1999pv,Berenstein:2000hy,Gomis:2000ej,Klein:2000tf,Sharpe:2000ki,Klein:2000qw,Sharpe:2000wu,Gaberdiel:2000fe,CG,Hauer:2001tp,Gaberdiel:2004vx}.
D-branes in theories with discrete torsion often give rise to
projective representations of the orbifold group on the Chan-Paton
factors. However, this feature does not describe discrete torsion
branes uniquely, as projective representations can also appear on
orbifolds without discrete torsion \cite{CG}. In this note we want to
address the following open problem of D-branes in discrete torsion
orbifolds. If we consider a $\mZ_n\times\mZ_n$ orbifold with discrete
torsion, we can calculate the RR potentials in the theory. Since
D-branes act as sources of RR flux, they have to couple to these RR
potentials. In the case of a $\mZ_2\times\mZ_2$ orbifold, BPS D-branes
have been constructed which couple to all RR potentials. The branes
which couple to the twisted RR potentials were given by fractional
BPS-branes with one Dirichlet and one Neumann boundary condition along
each complex direction in the orbifold \cite{Gaberdiel:2000fe}.

However, for the case $n$ odd the only known D-branes which couple to
the twisted RR potentials are non-BPS branes \cite{CG}. This is
somewhat unusual, since one should expect the
branes which carry the minimal charges to be BPS. On a technical
level the problem seemed to be that the BPS D-brane which carried the
twisted charges in the $\mZ_2\times\mZ_2$ case was not compatible with
the orbifold action for odd $n$.

The aim of this note is to construct a BPS brane which couples to
twisted RR charges in a $\mZ_3\times\mZ_3$ orbifold with discrete
torsion by using the boundary state formalism (for reviews see
\cite{Diaconescu:1999dt,Gaberdiel:2000jr,Billo:2000yb,GaberdielDbranes}).
In order to construct D-branes which are BPS in this background, we
have to consider more general boundary conditions than the ones
previously considered. Namely, we impose permuted gluing conditions
that couple left moving fields in direction $i$, and right moving
fields in direction $j$.  Consistency of these gluing conditions under
the orbifold group leaves us basically one additional combination of
fields which can be glued together. We will then show that this
combination leads to a consistent BPS brane.

It might seem natural to interpret the boundary states for this kind
of gluing conditions as permutation branes
\cite{RecknagelPermutation,Gaberdiel:2002jr} for the U(1) theory.
However, in contradistinction to permutation boundary states in $N=2$
minimal models, which were recently considered in \cite{IlkaMRG,
  Andreas}, the ``permuted'' gluing conditions in the U(1) case
actually turn out to describe D3-branes at angles. This is in a sense
similar to the result of \cite{Gaberdiel:2004nv} where it was shown
that all $N=2$ superconformal boundary states at $c=3$ can be
described in terms of usual Neumann and Dirichlet branes. We will come back
to that point in section \ref{sec:torsion}.

This paper is organised as follows. In section
\ref{sec:discrete-torsion} we will briefly review the effect of
discrete torsion on the construction of orbifolds. In section
\ref{sec:branes-orbifolds} we will then remind the reader of the usual
construction of D-branes in orbifold theories with discrete torsion,
and in the following sections we will construct the well known example
of the D0-brane in the $\mZ_3\times \mZ_3$ orbifold with and without discrete
torsion. In section \ref{sec:torsion} we will construct a
BPS-brane in the $\mZ_3\times \mZ_3$ orbifold with discrete torsion and
show that it couples to the twisted RR potentials. Section
\ref{sec:conclusions} contains our conclusions. We have summarised our
conventions on RR zero modes in an appendix.

\section{Discrete torsion}
\label{sec:discrete-torsion}

Shortly after the introduction of orbifolds, Vafa \cite{Vafa:1986wx}
pointed out the possibility of discrete torsion orbifolds. Let us
briefly recall their definition.

Consider the orbifold of a manifold $\mathcal{M}$ by a discrete
Abelian group $\Gamma$. As is well known
\cite{DHVW}, the orbifold theory consists of the
$\Gamma$ invariant states of the original theory, and additional
twisted sectors that describe strings which are closed only up 
to a group transformation by an element of $\Gamma$. These are
strings which are closed only in $\mathcal{M}/\Gamma$ but not in
$\mathcal{M}$. In an Abelian orbifold, we have one twisted sector
$\mH_h$ for each element $h\in\Gamma$. Each twisted sector has again
to be projected onto the $\Gamma$ invariant states. The projector onto
group invariant states is given by
\begin{equation}
  \label{eq:1}
P_\Gamma =  \frac{1}{|\Gamma|}\sum_{g\in \Gamma}g.
\end{equation}
Therefore, the partition function of the theory is given by
\begin{equation}
  \label{eq:7}
  Z(q,\bar{q}) =   \frac{1}{|\Gamma|}\sum_{g,h\in \Gamma}Z(q,\bar{q}; g,h),
\end{equation}
where
\begin{equation}
  \label{eq:8}
  Z(q,\bar{q}; g,h) = \mTr_{\mH_h}(gq^{L_0}\bar{q}^{\tilde{L}_0}).
\end{equation}

The discrete torsion theory is characterised by the fact that the
partition function is modified to 
\begin{equation}
  \label{eq:t16}
  Z(q,\bar{q}) =   \frac{1}{|\Gamma|}\sum_{g,h\in
  \Gamma}\epsilon(g,h)Z(q,\bar{q}; g,h), 
\end{equation}
where $\epsilon(g,h)$ are phases which have to obey the additional
consistency requirements \cite{Vafa:1986wx}
\begin{align}
  \label{eq:t7}
  \epsilon(g_1 g_2, g_3) &= \epsilon(g_2, g_3)\epsilon(g_1, g_3)\,, \\
  \label{eq:t8}
  \epsilon(g, h) &= \epsilon(h,g)^{-1}\,,\\
  \label{eq:t9}
  \epsilon(g, g)&=1\,.
\end{align}
The distinct discrete torsion phases are in one-to-one correspondence
with the second group cohomology H$^2$($\Gamma$,U(1)). 

We can interpret the appropriate projector for the theory with
discrete torsion to be given by
\begin{equation}
  \label{eq:18}
  P_\Gamma|_{\mH_h} = \frac{1}{|\Gamma|} \sum_{g\in \Gamma} \epsilon(g,h) g|_{\mH_h}\,.
\end{equation}
Alternatively, we can interpret the theory with discrete torsion as
the one where $g\in \Gamma$ acts on the sector $\mH_h$ as 
\begin{equation}
  \label{eq:16}
  \hat{g}|_{\mH_h} = \epsilon(g,h) g|_{\mH_h}.
\end{equation}
Due to the representation property of the discrete torsion phases, this
gives a well-defined action of $\Gamma$ on $\mH_h$. From the conformal
field theory point of view there is thus no fundamental difference
between the theory with discrete torsion and the theory without. The
only difference lies in the definition of the action of the orbifold
group on the twisted sectors which is \emph{a priori} not well
defined. While the action of the orbifold group in the untwisted
sector is fixed by the geometrical action of the group, this is not
true for the twisted sectors. 

\section{D-branes on orbifolds with discrete torsion}
\label{sec:branes-orbifolds}

In this section we want to spell out some properties of D-branes on
orbifolds with discrete torsion. There exists quite an extensive
literature on these matters, see for example
\cite{Gimon:1996rq,Quivers,Douglas:1998xa,Douglas:1999hq,Diaconescu:1999dt,Gaberdiel:1999ch,Billo:2000yb,CG,Gaberdiel:2000fe,Gomis:2000ej,Stefanski:2000fp,Bergman:2000ni
}.  Let us assume that spacetime is a product of Minkowski space and
an orbifold. Let us first consider a bulk D-brane at a generic point
of the orbifold.  If we start with a D-brane in the covering space, we
have to add images of this brane, in order to obtain an orbifold
invariant configuration. As we are looking at a generic point of the
orbifold, we will need $|\Gamma|$ copies of the original brane. The
$\Gamma$-invariant open strings that end on any two of these branes
describe the excitations of the 
D-branes. The action of $\Gamma$ on an open string state $\ket{\psi,
  ij}$ can be decomposed into an action on the oscillators, and an
action on the Chan-Paton factors
\begin{equation}
  \label{eq:20}
  g\ket{\psi, ij} = \gamma(g)_{ii'}\ket{U(g)\psi, i'j'}\gamma(g)^{-1}_{j'j}.
\end{equation}
In the case of a bulk brane the action on the Chan-Paton factors is
given by the regular representation of $\Gamma$ which has indeed
dimension $|\Gamma|$.

If we look at a D-brane near a fixed point of the orbifold, the
dimension of the representation $\gamma$ may be smaller. This is a
consequence of the 
fact that fewer preimages of the brane are needed at a fixed
point. Branes for which the dimension of $\gamma$ is strictly smaller
than $|\Gamma|$ are called \emph{fractional} D-branes. Fractional
branes are confined to reside on the fixed points because they can not
move off into the bulk. This is simply a consequence of the fact that
the dimension of $\gamma$ for those branes is smaller than for
ordinary bulk branes. To obtain a bulk brane from fractional branes,
one has to combine several fractional branes. 

By adding D-branes to a given string theory we introduce open strings
into a consistent closed string theory. In order to analyse the
consistency conditions for the open string sector, it is often useful
to compute the cylinder diagram between two (possibly identical)
D-branes.  This cylinder amplitude has two rather different
interpretations, depending on which world-sheet coordinate is given the
r\^ole of the time coordinate. In the open-string picture this diagram
is a one-loop vacuum diagram.  In the closed string picture 
it describes the tree-level exchange of closed strings between
two D-branes. The condition that the
closed and the open string descriptions should match imposes strong
constraints on the possible open string sectors which can be added
consistently to a given closed string theory.

In the closed string picture each D-brane can be
described as a boundary state $\Bket{D}$ which is a coherent state in
the closed string Hilbert space.
The distinction between bulk branes and fractional branes can as well
be seen in the boundary state approach. The boundary state of a bulk
brane has only components in the untwisted string Hilbert spaces,
whereas boundary states which correspond to fractional branes have
components in at least one twisted sector, as well. Components of a
D-brane in a given sector describe the coupling of the brane to closed
strings in that sector. The contribution of the $h$-twisted sector to
the cylinder diagram corresponds in the open string description to an
insertion of $h$ in the one-loop diagram. The sum over twisted sectors
then reproduces the projection operator \eqref{eq:1}. This implies
that boundary states with a non-trivial contribution in the
$h$-twisted sector lead to open strings for which the annulus diagram
contains non-trivial contributions in the sector with $h$
insertion. 

\section{The $\mZ_3\times\mZ_3$ orbifold with discrete torsion}
\label{sec:z_3-orbi}

We will consider the example of a $\mZ_3\times\mZ_3$ orbifold with
discrete torsion of a Type II theory on $\mC^3$. It is useful to
introduce complexified coordinates along the orbifold directions, as the
orbifold group then has a natural diagonal action on the coordinate
fields. The generators of the orbifold group $g_1$ and $g_2$ act as
\begin{subequations}
  \label{eq:17}
\begin{align}
  g_1: (z^1, z^2, z^3) &\mapsto (z^1, e^{-2 \pi i/3} z^2, e^{2 \pi i/3}
  z^3) \,,\\
  g_2: (z^1, z^2, z^3) &\mapsto ( e^{2 \pi i/3} z^1, z^2,
  e^{-2 \pi i/3} z^3)\,.
\end{align}
\end{subequations}

The different possible choices of discrete torsion phases are in
one-to-one correspondence with the second cohomology class
H$^2$($\Gamma$,U(1)). It is known that for $\Gamma=\mZ_m\times\mZ_m$,
H$^2$($\Gamma$,U(1)) is isomorphic to $\mZ_m$. Thus in our example the
group of different discrete torsion phases is isomorphic to
$\mZ_3$. We therefore have the possibility to set
\begin{equation}
  \label{eq:19}
  \omega=\epsilon(g_1,g_2)=
  \begin{cases}
    1\,, \\
    e^{2 \pi i /3}\,, \\
    e^{-2 \pi i /3}\,. 
  \end{cases}
\end{equation}
The first choice corresponds to the case which is usually called the
theory without discrete torsion. It should be clear that for
$\Gamma=\mZ_3\times \mZ_3$ fixing $\epsilon(g_1,g_2)$ determines all
discrete torsion phases. This is due to the consistency conditions
mentioned in section \ref{sec:discrete-torsion}.

In order to summarise the RR ground state spectrum, let us state the
Hodge diamond for the $\mZ_3\times\mZ_3$ orbifold. The untwisted sector
contributes
\begin{equation}
  \label{eq:23}
  \begin{matrix}
    & & &1& & & \\
    & &0& &0& & \\
    &0& &3& &0& \\
   1& &0& &0& &1\\
    &0& &3& &0& \\
    & &0& &0& & \\
    & & &1 & & & \\
  \end{matrix}
\end{equation}
to the Hodge diamond  \cite{Douglas:1999hq}. In the theory with
$\omega=1$, the contribution from the twisted sectors is
given by
\begin{equation}
  \label{eq:24}
  \begin{matrix}
    & & &0& & & \\
    & &0& &0& & \\
    &0& &7& &0& \\
   0& &0& &0& &0\\
    &0& &7& &0& \\
    & &0& &0& & \\
    & & &0& & & \\
  \end{matrix}
\end{equation}
For $\omega \neq 1$ we obtain on the other hand
\begin{equation}
  \label{eq:25}
  \begin{matrix}
    & & &0& & & \\
    & &0& &0& & \\
    &0& &0& &0& \\
   0& &3& &3& &0\\
    &0& &0& &0& \\
    & &0& &0& & \\
    & & &0& & & \\
  \end{matrix}
\end{equation}
In the theory with $\omega\neq 1$,
the BPS branes constructed in the literature
\cite{Douglas:1999hq,Diaconescu:1999dt} do not couple to these twisted
RR potentials. The only D-brane hitherto constructed which couples to
the above twisted RR potentials for $\omega\neq 1$ is non-BPS
\cite{CG}, and does not 
couple to the untwisted RR potentials. In the sequel we will  
construct a BPS D-brane which couples to all of these potentials.

In order to construct this brane, we will first fix our conventions by
constructing a fractional $D(r,0)$-brane in the theory with
$\omega=1$.  We denote by a $D(r,s)$-brane a $Dp$-brane with $r$
Neumann boundary conditions in the directions unaffected by the
orbifold, and $s$ Neumann directions along the orbifold, where
$p=r+s$.  This brane has been constructed before in
\cite{Diaconescu:1999dt} and we will repeat the analysis here, mainly
to fix some conventions and phases. The fractional $D(r,0)$-brane
couples to all twisted sectors.

\section{The $D(r,0)$-brane}
\label{sec:d0-brane}

\subsection{The untwisted sector}
\label{sec:construction-d0}

We will now construct the boundary state of the $D(r,0)$-brane with
$\omega=\epsilon(g_1,g_2)=1$. This section 
is basically a review of the work in \cite{Diaconescu:1999dt}, and we
need it to introduce some notation and conventions. 

The $D(r,0)$-brane has Dirichlet gluing conditions in all spatial
directions along the orbifold and $r$ Neumann gluing conditions in the
other directions. With this notation it is possible to treat type IIA
and type IIB theories on an equal footing. The number of Neumann
conditions along the directions unaffected by the orbifold can then be
chosen in accordance with the GSO projection \cite{Gaberdiel:1999ch}.

For concreteness we will consider the directions
$2, \dots, 7$ to be the directions along which the orbifold acts. The
gluing conditions are given by
\begin{equation}
  (a^i_{m} - \tilde{a}^i_{-m})\Iket{D(r,0)} = 0\,, \qquad i=2,\dots,7\,.
\end{equation}

For the sake of simplicity, we will work in light cone gauge by taking
$x^8\pm x^9$ as light-cone coordinates, after a double Wick rotation on
$x^0$ and $x^8$ \cite{Green:1996um}.  As has been mentioned in the
previous chapter, it is convenient to group the directions which are
affected by the orbifold action into complex pairs.

The bosonic fields of the closed string are given by
\[
X^\mu  = x^\mu + 2\pi p^\mu t + \frac{i}{\sqrt{2}}
\sum_{n\neq0}\frac{1}{n}(a_n^\mu e^{-2\pi i n (t-\sigma)} +
\tilde{a}_n^\mu e^{-2\pi i n (t+\sigma)})\,,
\]
and the corresponding fermionic fields are given by
\begin{align*}
  \lambda^\mu &= \sqrt{2\pi} \sum_r \lambda^\mu_r e^{-2\pi i r
  (t-\sigma)} \,,\\ 
  \tlambda^\mu &= \sqrt{2\pi} \sum_r \tlambda^\mu_r e^{-2\pi i r
  (t+\sigma)} \,. 
\end{align*}
If we group these into complex pairs, we obtain 
\begin{align*}
  Z^i&:=\frac{1}{\sqrt{2}}(X^{2i} + iX^{2i+1})\,, &
  Z^{\ibar}&:=\frac{1}{\sqrt{2}}(X^{2i} - iX^{2i+1}) \,, \qquad
  i=1,2,3 \\
  \psi^i&:=\frac{1}{\sqrt{2}}(\lambda^{2i} + i\lambda^{2i+1})\,, &
  \psi^{\ibar}&:=\frac{1}{\sqrt{2}}(\lambda^{2i} - i\lambda^{2i+1})\,.
\end{align*}
On the fields $Z^i$ ($\psi^i$) the
orbifold group acts as given in equation \eqref{eq:17}, and on the
$Z^{\ibar}$ ($\psi^{\ibar}$) it acts with the opposite phases.

On the level of the oscillators we obtain 
\begin{align*}
  \alpha^i_n&= \frac{1}{\sqrt{2}}(a^{2i}_n + ia^{2i+1}_n)\,,
  \qquad i=1,2,3 \\
  \alpha^{\ibar}_n&= \frac{1}{\sqrt{2}}(a^{2i}_n -
  ia^{2i+1}_n) \,, 
\end{align*}
where we adopt the same conventions for the right-movers, and
analogously for the fermionic modes we write
\begin{align*}
  \psi^i_r&=\frac{1}{\sqrt{2}}(\lambda^{2i}_r + i\lambda^{2i+1}_r)\,,
   \\ 
  \psi^{\ibar}_r&= \frac{1}{\sqrt{2}}(\lambda^{2i}_r -
  i\lambda^{2i+1}_r) \,.
\end{align*}
The (anti-)commutation relations for the complexified oscillators are
given by 
\begin{align*}
  [\alpha^{i}_n,\alpha^{\bar{\jmath}}_m]
  &=[\tilde{\alpha}^{i}_n,\tilde{\alpha}^{\bar{\jmath}}_m] =
  n\delta_{n+m}\delta^{ij}\,,  \\
  \{\psi^i_r,\psi_s^{\bar{\jmath}}\} &= \{\tpsi^i_r,\tpsi_s^{\bar{\jmath}}\} =
  \delta_{r+s}\delta^{ij}\,.
\end{align*}
In terms of complexified coordinates along the orbifold, the gluing
conditions are given by
\begin{subequations}
\label{eq:5}  
\begin{align}
  (\alpha_n^i - \tilde{\alpha}_{-n}^i)\Iket{D(r,0)}&=0 \,,\\
  (\alpha_n^{\ibar} -\tilde{\alpha}_{-n}^{\ibar})\Iket{D(r,0)}&=0 \,, \\
  (\psi_r^i - i \eta \tilde{\psi}_{-r}^i)\Iket{D(r,0),\eta}&=0 \,,\\
  (\psi_r^{\ibar} - i \eta \tilde{\psi}_{-r}^{\ibar})\Iket{D(r,0),\eta}&=0\,.
\end{align}
\end{subequations}
Here $\eta=\pm1$ and denotes the possible spin structures on the world-sheet
\cite{GaberdielDbranes}. 

The NS-NS and RR Ishibashi states which solve the above gluing
conditions are the coherent states
\begin{multline}
  \label{eq:4}
  \Iket{D(r,0),\eta}_{\aob{NSNS}{RR}} = \Iket{Dr}\exp\Bigg[\sum_{n=1}^\infty
      \frac{1}{n}\left( \sum_{i=1}^3
        \alpha^i_{-n}\tilde{\alpha}^{\ibar}_{-n}  +
        \alpha^{\ibar}_{-n}\tilde{\alpha}^i_{-n}\right) + \\
      +i\eta\sum_{r>0}\sum_{i=1}^3
        \psi^i_{-r}\tilde{\psi}^{\ibar}_{-r} + 
        \psi^{\ibar}_{-r}\tilde{\psi}^i_{-r}
      \Bigg]\ket{0,\eta}_{\aob{NSNS}{RR}}\,.
\end{multline}
We have denoted by $\Iket{Dr}$ the components of the coherent
state in the directions which were not orbifolded.  In the NS-NS
sector $\ket{0,\eta}_{NSNS}$ is the unique NS-NS vacuum, whereas in
the RR sector we have to work a little harder. The RR ground states have to
solve \eqref{eq:5} for the fermionic zero modes.  Let us define the
$\psi^i_0$s to be raising operators, and the $\psi^{\ibar}_0$s to be
lowering operators. We denote the state which is annihilated by
$\psi^{\ibar}_0$ with $\ket{-}$ and the state
$\psi^i_0\ket{-}=\ket{+}$. We will adopt the same definition for the
right-movers, as well. The ordering conventions for the fermionic
zero modes are given in the appendix. Gluing conditions of the form
\eqref{eq:5} are solved by the state $\ket{+}\otimes\ket{-} -
i\eta\ket{-}\otimes\ket{+}$. The next step is to build the tensor
product of the ground states of the various directions.  The ground
state which satisfies all gluing conditions is thus given by
\begin{equation}
  \label{eq:49}
  \prod_{i=1}^3(\psi^i_0 - i\eta\tilde{\psi}^i_0)\ket{0}_{RR}\,,\qquad
  \text{with}\qquad \ket{0}_{RR}= \ket{---}\otimes\ket{---}\,.
\end{equation}
Here we only considered the gluing conditions and the RR ground state
in the orbifold directions. For the other directions, we can simply
tensor the appropriate RR ground state with the above state, once we
have decided on the gluing conditions.

We still have to impose the GSO projection. In the NS-NS sector the
vacuum is odd under $(-1)^F$ and $(-1)^{\tilde{F}}$ and therefore we
obtain
\[
(-1)^F\Iket{D(r,0),\eta}_{NSNS} = (-1)^{\tilde{F}}\Iket{D(r,0),\eta}_{NSNS}
=-\Iket{D(r,0),-\eta}_{NSNS}\,.  
\]
In the RR sector the GSO operators act as the chirality operators on
the ground state, and for a general D$p$-brane we define
$\Iket{Dp,-\eta}_{RR}$ by
\begin{equation}
(-1)^F\Iket{Dp,\eta}_{RR} :=  \Iket{Dp,-\eta}_{RR}\,.
\end{equation}
This fixes the constant of proportionality. Then the action of
$(-1)^{\tilde{F}}$ on this state is given by
\begin{equation}
(-1)^{\tilde{F}}\Iket{Dp,\eta}_{RR} = (-1)^{p+1}\Iket{Dp,-\eta}_{RR}\,,
\end{equation}
where $p=r+s$.
Therefore the GSO invariant combinations are given by 
\[\Bket{D(r,0)}_{NSNS}
=\frac{1}{\sqrt{2}}\big(\Iket{D(r,0),+}_{NSNS}-\Iket{D(r,0),-}_{NSNS}\big)
\]
for the NS-NS sector, and for the RR sector of the $D(r,0)$-brane we obtain
\begin{equation}
  \label{eq:RR-boundary}
  \Bket{D(r,0)}_{RR}
  =\frac{1}{\sqrt{2}}\big(\Iket{D(r,0),+}_{RR}+\Iket{D(r,0),-}_{RR}\big)\,.
\end{equation}
\eqref{eq:RR-boundary} is  invariant under the GSO projection of the type IIA
theory if $r$ is even, and under the GSO projection of the type IIB
theory if $r$ is odd.

In the next step let us see how the boundary state \eqref{eq:4}
behaves under the action of the orbifold group. The oscillators are
mapped according to \eqref{eq:17}  to
\begin{subequations}
  \label{eq:9}
\begin{align}
  g_k \alpha_n^i g_k^{-1} &= e^{2 \pi i \nu^i_k}\alpha_n^i  \,,
  & g_k\alpha_n^{\ibar} g_k^{-1}&= e^{-2 \pi i \nu^i_k} \alpha_n^{\ibar}, \\
    g_k \psi_n^i g_k^{-1} &= e^{2 \pi i \nu^i_k}\psi_n^i  \,,
  & g_k\psi_n^{\ibar} g_k^{-1} &= e^{-2 \pi i \nu^i_k} \psi_n^{\ibar}, 
\end{align}
\end{subequations}
where $\nu_1 = (0, -1/3, 1/3)$ and $\nu_2 = (1/3, 0, -1/3)$, and
similarly for the right-movers.  Therefore the terms in the
exponential of \eqref{eq:4} are invariant under the group action.
Since the orbifold action on the untwisted NS-NS vacuum is trivial, we
only have to determine the action of the orbifold group on the RR
ground state. The action on the RR ground state is given by the action
on the oscillators and the action on $\ket{0}_{RR}$. It is easy to see
that the combinations of oscillators which appear if we expand
\eqref{eq:49} are invariant under $\Gamma$.  The orbifold action maps
annihilation operators again to annihilation operators, therefore a
state which has been annihilated by all annihilators will still be
destroyed by all annihilation operators after the action of the
orbifold group. Therefore $\ket{0}_{RR}$ must carry some
representation of the orbifold group,
\[ g_k^n \ket{0}_{RR} = \rho^n_k\ket{0}_{RR}\,.\]
The only possible values for $\rho$ are third roots of unity.
We know on the
other hand that the RR ground state of the $D(r,0)$-brane is not projected
out by the orbifold projection \cite{Diaconescu:1999dt} and therefore
we can fix $\rho_k=1$.

\subsection{The twisted sector}
\label{sec:twisted-sector}

Let us now turn to the twisted sectors. In the twisted sectors the
oscillator moding changes since the string only closes up to a group
transformation.  For the orbifold action in question, the mode
expansions for the sector twisted by $g_k^m$ are given by
\begin{align*}
  Z^i &= \frac{i}{\sqrt{2}}\left(\sum_{r \in \mZ + m\nu_k^i} 
    \frac{1}{r} 
    \alpha^i_{r} e^{- 2 \pi i r(t-\sigma)} + 
    \sum_{r \in \mZ - m\nu_k^i} 
    \frac{1}{r} \tilde{\alpha}^i_{r} e^{- 2 \pi i r(t+\sigma)} \right) \,,\\
  Z^{\ibar} &= \frac{i}{\sqrt{2}}\left(\sum_{r \in \mZ - m\nu_k^i} 
    \frac{1}{r} 
    \alpha^{\ibar}_{r} e^{- 2 \pi i r (t-\sigma)} + 
    \sum_{r \in \mZ + m\nu_k^i} 
    \frac{1}{r} \tilde{\alpha}^{\ibar}_{r} e^{- 2 \pi i
    r(t+\sigma)} \right) \,,\\
  \psi^i &= \sqrt{2\pi}\sum_{r \in \mZ + m\nu_k^i + s}  \psi^i_{r} e^{- 2 \pi
    ir(t-\sigma)}\,, \\
  \psi^{\ibar} &= \sqrt{2\pi}\sum_{r \in \mZ - m\nu_k^i + s}
    \psi^{\ibar}_{r}  e^{- 2 \pi i r(t-\sigma)} \,,\\
  \tpsi^i &= \sqrt{2\pi}\sum_{r \in \mZ - m\nu_k^i + s} \tpsi^i_{r} 
  e^{- 2 \pi i r (t+\sigma)} \,,\\
  \tpsi^{\ibar} &= \sqrt{2\pi}\sum_{r \in \mZ  + m\nu_k^i + s}
    \tpsi^{\ibar}_{r} e^{- 2 \pi i r(t+\sigma)} \,,
\end{align*}
where the $\nu_k^i$ are as defined below equation \eqref{eq:9}, and
$s=\tfrac{1}{2}$ ($s=0$) in the NS (R) sector. 
Note that the mode shifts of the right-movers are opposite to the mode
shifts of the left-movers. This is due to the fact that the mode
shifts are determined by the behaviour under the transformation
$\sigma\rightarrow \sigma+1$ which differs by a minus sign between
left-movers and right-movers.
The
(anti-)commutation relations for the twisted oscillators are given by
\begin{align}
  \label{eq:10}
  [\alpha^i_{n+\nu}, \alpha^{\jbar}_{m-\nu}] =
  [\tilde{\alpha}^i_{n+\nu}, \tilde{\alpha}^{\jbar}_{m-\nu}]
  &=(n+\nu)\delta_{n+m}\,\delta^{ij}, \\
  \{\psi^i_{r+\nu}, \psi^{\jbar}_{s-\nu}\} = 
  \{\tpsi^i_{r+\nu}, \tpsi^{\jbar}_{s-\nu}\} &= \delta_{r+s}\,\delta^{ij},
\end{align}
and the other relations vanish. The gluing conditions in the 
$g^m_k$-twisted 
sector are 
those given by \eqref{eq:5} with the shifted modes in the orbifold
directions
\begin{subequations}
\label{eq:11}
\begin{align}
   (\alpha_{n+m\nu_k^i}^i - \tilde{\alpha}_{-n-m\nu_k^i}^i)\Iket{D(r,0)}&=0 \,,\\
  (\alpha_{n-m\nu_k^i}^{\ibar}
  -\tilde{\alpha}_{-n+m\nu_k^i}^{\ibar})\Iket{D(r,0)}&=0 \,,\\ 
  (\psi_{r+m\nu_k^i}^i - i \eta
  \tilde{\psi}_{-r-m\nu_k^i}^i)\Iket{D(r,0),\eta}&=0 \,, \\ 
  (\psi_{r-m\nu_k^i}^{\ibar} - i \eta
  \tilde{\psi}_{-r+m\nu_k^i}^{\ibar})\Iket{D(r,0),\eta}&=0\,.   
\end{align}
\end{subequations}
Therefore the Ishibashi states in this sector are given by the coherent states 
\begin{multline}
\label{eq:12}
  \Iket{D(r,0),\eta}_{\aob{NSNS}{RR},T_{g_k^m}} = \\
  \Iket{Dr}\exp\Bigg[\sum_{n}
       \sum_{i=1}^3\left( \frac{1}{n-m\nu_k^i}
      \alpha^i_{-n+m\nu_k^i}\tilde{\alpha}^{\ibar}_{-n+m\nu_k^i}  +  
        \frac{1}{n+m\nu_k^i}
      \alpha^{\ibar}_{-n-m\nu_k^i}\tilde{\alpha}^i_{-n-m\nu_k^i}\right)
      + \\  
      +i\eta\sum_{r}\sum_{i=1}^3\bigg(
        \psi^i_{-r+m\nu_k^i}\tilde{\psi}^{\ibar}_{-r+m\nu_k^i} + 
        \psi^{\ibar}_{-r-m\nu_k^i}\tilde{\psi}^i_{-r-m\nu_k^i}\bigg)
      \Bigg]\ket{0,\eta}_{\aob{NSNS}{RR},T_{g_k^m}}\,,
\end{multline}
where the summation over $n$ and $r$ should start appropriately, such
that all creation operators appear in the exponential.
For simplicity, and in an attempt to clutter the notation not any
further, we have only written down a sector twisted by
$g_k^m$. There are of course also sectors in the complete
theory which are twisted by $g_1^mg_2^l$ and we thus obtain a total of
eight twisted sectors. In the additional twisted sectors the reasoning
is identical, and we have not explicitly written them down here.

Now we have to turn our attention to the Ramond ground state. In
the $g_1$-twisted sector we have zero modes in the complex 1-plane,
and in the directions unaffected by the orbifold. We have again
to determine the action of the orbifold group on the relevant zero
modes. The gluing conditions for the zero modes are as before, and
therefore the state which solves the gluing conditions is given by
\begin{equation}
  \label{eq:13}
  \ket{0,\eta}_{RR,T_{g_1^m}} = (\psi^1_0 - i\eta
  \tilde{\psi}^1_0)\ket{-}\otimes\ket{-}\,.
\end{equation}
As before, we have to determine the action of $g_2$ on this ground
state. (The action of $g_1$ is of course trivial in the 1-direction.)
We can do so as follows: the action of $g_2$ on
$\ket{0,\eta}_{RR,T_{g_1^m}}$ is given by 
\[
g_2^n \ket{0,\eta}_{RR,T_{g_1^m}} = \omega^{mn}\zeta^n \rho_1^n
\ket{0,\eta}_{RR,T_{g_1^m}} \,,
\] 
where $\zeta=e^{2 \pi i/3}$ and we have parametrised the action of
$g_2$ on $\ket{-}\otimes\ket{-}$ by $\rho_1$.  As before, we know from
previous work that this state is not projected out in the theory with
$\omega=1$. Therefore, we have to choose $\rho_1=\zetabar$. The
construction for the $g_2$ and $g_1g_2$ twisted sectors is identical.
The $g_1g_2^2$-twisted sector does not possess any zero modes along
the orbifold directions.  As before in the untwisted sector, we have
to combine Ishibashi states with different values of $\eta$ in order
to obtain a GSO invariant boundary state. The complete boundary state
of the fractional $D(r,0)$-brane is thus a sum of the boundary states
of the various twisted sectors,
\begin{equation}
  \label{eq:14}
  \Bket{D(r,0)} = \frac{1}{3}\sN\sum_{m,n=0}^2 \Bket{D(r,0)}_{T_{g_1^mg_2^n}}\,.
\end{equation}
In order to fix the normalisation, the next step would  be  to compute
the overlap for this brane and compare it with the open string
result. This has been done very explicitly in
\cite{Diaconescu:1999dt} and we will not repeat this calculation here.

\section{$\mC^3/\mZ_3\times\mZ_3$ with discrete torsion}
\label{sec:torsion}

Let us now consider the same orbifold with $\omega\neq 1$. Since
discrete torsion only alters the group action in the twisted sectors,
the analysis of the untwisted sector remains unchanged. However, as
can easily be seen 
from the discussion of the twisted sector ground states, the RR ground
states of the $D(r,0)$-brane will be projected out if we introduce
additional discrete torsion phases. Therefore the $D(r,0)$-brane does
not couple to the twisted RR potentials.  The Hodge diamond
\eqref{eq:25} shows that there exist twisted RR potentials in
the theory with $\omega\neq 1$. Since the $D(r,0)$-brane does not
couple to them, we have to construct some other brane which is charged
under these twisted RR potentials.  In \cite{CG} it was shown that it
is possible to construct non-BPS branes with one Neumann direction
along the orbifold, say the 1-direction, and with components only in
the $g_1$-twisted sectors that couple to the twisted RR potentials.
This brane however does not couple to any untwisted RR potentials.

\subsection{The untwisted sector}
\label{sec:untwisted-sector}

We want to consider more general gluing conditions than the ones which
have been considered in previous works. By doing so, we will be able
to construct a BPS-brane coupling to twisted and untwisted RR
potentials.  Namely, we want to consider gluing conditions where we
impose one Neumann and one Dirichlet condition along the complex
1-direction as has been done before, but we want to permute the gluing
in the 2 and 3-directions. In the end, we want to keep Ishibashi states
in the $g_1$-twisted sector, therefore we have to make sure that the
gluing conditions imposed can be satisfied by $g_1$-twisted
oscillators. Considering the action of $g_1$ along the 2 and 3
direction thus leaves us only with one new choice of gluing
conditions, namely we can impose
\begin{subequations}
  \label{eq:15}   
  \begin{align}
    (\alpha^1_{r} + \tilde{\alpha}^{\bar{1}}_{-r})\Bket{B} &=0 \,, \\
    (\alpha^{\bar{1}}_{r} +  \tilde{\alpha}^{1}_{-r})\Bket{B} &=0 \,, \\
    (\alpha^2_{r} + \tilde{\alpha}^{\bar{3}}_{-r})\Bket{B} &=0 \,,
    \label{eq:pair1}  \\ 
    (\alpha^{\bar{2}}_{r} +  \tilde{\alpha}^{3}_{-r})\Bket{B} &=0
    \label{eq:pair2}\,, \\ 
    (\alpha^3_{r} + \tilde{\alpha}^{\bar{2}}_{-r})\Bket{B} &=0 \,, \\
    (\alpha^{\bar{3}}_{r} +  \tilde{\alpha}^{2}_{-r})\Bket{B} &=0 \,,
  \end{align}
\end{subequations}
for the bosons and analogously
\begin{subequations}
  \label{eq:21}
  \begin{align}
    (\psi^1_{r} + i\eta \tilde{\psi}^{\bar{1}}_{-r})\Bket{B} &=0 \,, \\
    (\psi^{\bar{1}}_{r} + i\eta \tilde{\psi}^{1}_{-r})\Bket{B} &=0 \,, \\
    (\psi^2_{r} + i\eta \tilde{\psi}^{\bar{3}}_{-r})\Bket{B} &=0 \,, \\
    (\psi^{\bar{2}}_{r} + i\eta \tilde{\psi}^{3}_{-r})\Bket{B} &=0 \,, \\
    (\psi^3_{r} + i\eta \tilde{\psi}^{\bar{2}}_{-r})\Bket{B} &=0 \,, \\
    (\psi^{\bar{3}}_{r} + i\eta \tilde{\psi}^{2}_{-r})\Bket{B} &=0 \,,    
  \end{align}
\end{subequations}
for the fermions along the orbifold directions. The directions
unaffected by the orbifold remain as before. 
We could replace pairs of plus signs with minus signs in the gluing
conditions  \eqref{eq:15} and \eqref{eq:21}. For example we could
instead of \eqref{eq:pair1} and 
\eqref{eq:pair2} impose 
\begin{align*}
      (\alpha^2_{r} - \tilde{\alpha}^{\bar{3}}_{-r})\Bket{B} &=0 \,,
\ \qquad \text{and}\  &
    (\alpha^{\bar{2}}_{r} -  \tilde{\alpha}^{3}_{-r})\Bket{B} &=0
    \,.
\end{align*}
However, this choice only determines which of the real directions are
glued together with Neumann, and which with Dirichlet boundary
conditions. If we try to change only one of the signs, the gluing
conditions cease to commute.

If we compare the gluing conditions \eqref{eq:15} and \eqref{eq:21}
with the action of $g_1$ on the 
oscillators, we see that they are compatible with twisting by $g_1$.
The action of $g_2$ introduces phases on these gluing
conditions. Therefore, in order to construct a brane which
could possibly be invariant under the orbifold group with the above
gluing conditions, the following system of gluing
conditions has to be considered.
\begin{subequations}
  \label{eq:26}
  \begin{align}
    (\alpha^1_{r} + \zeta^n\tilde{\alpha}^{\bar{1}}_{-r})\Iket{B_n} &=0 \,, \\
    (\alpha^{\bar{1}}_{r} +
    \zetabar^n\tilde{\alpha}^{1}_{-r})\Iket{B_n} &=0 \,,\\ 
    (\alpha^2_{r} + \zeta^n\tilde{\alpha}^{\bar{3}}_{-r})\Iket{B_n} &=0 \,, \\
    (\alpha^{\bar{2}}_{r} + \zetabar^n
    \tilde{\alpha}^{3}_{-r})\Iket{B_n} &=0 \,, \\
    (\alpha^3_{r} + \zeta^n\tilde{\alpha}^{\bar{2}}_{-r})\Iket{B_n} &=0 \,, \\
    (\alpha^{\bar{3}}_{r} +
    \zetabar^n\tilde{\alpha}^{2}_{-r})\Iket{B_n}&=0 \,.
  \end{align}
\end{subequations}
We impose analogous gluing conditions for the fermions, where we have
again set $\zeta=e^{2 \pi i /3}$, such that in particular $\zeta^3=1$.
With these gluing conditions we obtain three Ishibashi states
$\Iket{B_n}$ such that $g_2\Iket{B_n} = \Iket{B_{n+1}}$, where we
identify $\Iket{B_3}$ with $\Iket{B_0}$, more generally we take 
$n \mod{3}$. The brane in question is then the orbifold invariant sum of
these Ishibashi states.  

It is possible to use these gluing conditions to obtain an orbifold
invariant, supersymmetric boundary state, as we will see in the
following.  In the case at hand we find it convenient to use the name
permutation brane for the brane we are constructing, but this object
does not fall into a new class of D-branes.  This is fundamentally
different from the case of tensor products of $N=2$ minimal models for
example, where 
permutation branes are often new objects.

The Ishibashi states which solve the gluing 
equations \eqref{eq:15} and \eqref{eq:26} in the untwisted sector are
given by 
\begin{multline}
  \label{eq:22}
  \Iket{B_n}_{\aob{NSNS}{RR}} = \Iket{Dr}\exp\Bigg[
  -\sum_{k=1}^\infty
\frac{1}{k}\bigg(
        \zetabar^n\left(
          \alpha^1_{-k}\tilde{\alpha}^{1}_{-k} +
          \alpha^2_{-k}\tilde{\alpha}^{3}_{-k} +
          \alpha^3_{-k}\tilde{\alpha}^{2}_{-k} \right)\\ +
        \zeta^n \left(
          \alpha^{\bar{1}}_{-k}\tilde{\alpha}^{\bar{1}}_{-k} +
          \alpha^{\bar{2}}_{-k}\tilde{\alpha}^{\bar{3}}_{-k} +
          \alpha^{\bar{3}}_{-k}\tilde{\alpha}^{\bar{2}}_{-k} \right)
      \bigg) 
      -i\eta\sum_{r>0}\bigg(  
        \zetabar^n\left(
        \psi^1_{-r}\tilde{\psi}^{1}_{-r} + 
        \psi^2_{-r}\tilde{\psi}^{3}_{-r} + 
        \psi^3_{-r}\tilde{\psi}^{2}_{-r} \right)\\
      +\zeta^n \left(
        \psi^{\bar{1}}_{-r}\tilde{\psi}^{\bar{1}}_{-r} +
        \psi^{\bar{2}}_{-r}\tilde{\psi}^{\bar{3}}_{-r} +
        \psi^{\bar{3}}_{-r}\tilde{\psi}^{\bar{2}}_{-r}
      \right)
      \bigg)\Bigg]\ket{0,\eta}_{\aob{NSNS}{RR}}\,.
\end{multline}
One easily shows that $g_2\Iket{B_n} = \Iket{B_{n+1}}$ and $g_1
\Iket{B_n} = \Iket{B_n}$. 
The orbifold invariant combination is therefore given by 
\[
\Bket{B} = \sN \sum_{n=0}^2 \Iket{B_n}\,,
\]
where 
\[
\Iket{B_n} = \tfrac{1}{2}\left(\Iket{B_n,+}_{NSNS} - \Iket{B_n,-}_{NSNS} +
\epsilon(\Iket{B_n,+}_{RR} + \Iket{B_n,-}_{RR}\right) \,.
\]
In the previous equation $\epsilon$ is a sign and distinguishes
between branes and anti-branes. In order to simplify the equations we
will set $\epsilon=1$.  Let us now check if this brane does couple to
RR potentials.  To this end, we have to consider the Ramond ground
states of the brane system. Analogously to the $D(r,0)$-brane, we can
construct the ground states of these branes. In this case they are
given by
\begin{equation}
  \label{eq:29}
\Iket{B_n}^{(0)}_{RR} =  (\psi^1_0\tpsi^1_0
+ i\eta\zeta^n)(\psi^2_0\tpsi^3_0 + i\eta\zeta^n)(\psi^3_0\tpsi^2_0 +
i\eta\zeta^n)\ket{---}\otimes\ket{---}\,.
\end{equation}
If we expand these states out and take the sum over $n$, the ground
state of $\Bket{B}$ is given by
\begin{equation}
  \label{eq:30}
  \Bket{B}^{(0)}_{RR} = \ket{+++}\otimes\ket{+++} -i\eta
 \ket{---}\otimes\ket{---}\,,
\end{equation}
where we have, as before, only considered the RR ground states along
the orbifold.  Therefore, this D-brane does indeed couple to RR
potentials in the untwisted sector.

\subsection{The twisted sector}
\label{sec:twisted-sector-1}

Let us now turn to the construction of the twisted sector boundary
states. 
The construction of the $g_1$-twisted sector is basically identical to
the construction of the twisted sector for the $D(r,0)$-brane. The gluing
conditions are changed by the mode shifts, and the Ishibashi states in
the $g_1^m$-twisted sectors are then given by
\begin{multline}
  \label{eq:28}
  \Iket{B_n}_{\aob{NSNS}{RR},T_{g_1^m}} = \\ \Iket{Dr}\exp\Bigg[
  -\sum_{k}\bigg(
        \zetabar^n\left(\tfrac{1}{k}
          \alpha^1_{-k}\tilde{\alpha}^{1}_{-k} +
          \tfrac{1}{k+m/3}
          \alpha^2_{-k-m/3}\tilde{\alpha}^{3}_{-k-m/3} +
          \tfrac{1}{k-m/3}
          \alpha^3_{-k+m/3}\tilde{\alpha}^{2}_{-k+m/3} \right) +\\
        \zeta^n \left(
          \tfrac{1}{k}
          \alpha^{\bar{1}}_{-k}\tilde{\alpha}^{\bar{1}}_{-k} +
          \tfrac{1}{k-m/3}
          \alpha^{\bar{2}}_{-k+m/3}\tilde{\alpha}^{\bar{3}}_{-k+m/3} +
          \tfrac{1}{k+m/3}
          \alpha^{\bar{3}}_{-k-m/3}\tilde{\alpha}^{\bar{2}}_{-k-m/3}
  \right)\bigg)
     \\
      -i\eta\sum_{r}\bigg(
        \zetabar^n\left(
        \psi^1_{-r}\tilde{\psi}^{1}_{-r} + 
        \psi^2_{-r-m/3}\tilde{\psi}^{3}_{-r-m/3} + 
        \psi^3_{-r+m/3}\tilde{\psi}^{2}_{-r+m/3} \right) \\
      +\zeta^n \left(
        \psi^{\bar{1}}_{-r}\tilde{\psi}^{\bar{1}}_{-r} +
        \psi^{\bar{2}}_{-r+m/3}\tilde{\psi}^{\bar{3}}_{-r+m/3} +
        \psi^{\bar{3}}_{-r-m/3}\tilde{\psi}^{\bar{2}}_{-r-m/3}
      \right)
      \bigg)\Bigg]\ket{0,\eta}_{\aob{NSNS}{RR}}\,.
\end{multline}
Again, the summation should run over all creation operators.
The form of these Ishibashi states implies that 
\[
g_2^l \Iket{B_n}_{NSNS,T_{g_1^m}} =  \omega^{lm} \Iket{B_{n+l}}_{NSNS,T_{g_1^m}}\,.
\] 
We have incorporated discrete torsion by modifying
the action of $g_2$ on the ground state of the $g^m_1$-twisted sector.
It is now defined as 
\[
g_2^l \ket{0,\eta}_{NSNS,T_{g_1^m}} = \omega^{lm}
\ket{0,\eta}_{NSNS,T_{g_1^m}}\,.
\]
Therefore, an orbifold invariant combination of these Ishibashi states
is given by
\begin{equation}
  \label{eq:31}
  \Bket{B}_{T_{g_1^m}} = \sN \sum_{n=0}^2
  \omega^{mn}\Iket{B_n}_{T_{g_1^m}} \,.
\end{equation}
Let us now turn our attention to the RR ground states in the twisted
sector. The gluing conditions are given by
\begin{align*}
  (\psi^1_0 + i\eta\zeta^n\tpsi^{\bar{1}}_0)\Iket{B_n}_{T_{g_1}}^{(0)}
  &= 0 \,, \\
  (\psi^{\bar{1}}_0 +
  i\eta\zetabar^n\tpsi^{1}_0)\Iket{B_n}_{T_{g_1}}^{(0)} &= 0\,.
\end{align*}
We omit the gluing conditions in the directions unaffected
by the orbifold, as they are irrelevant for this problem.
These gluing conditions can be solved by states
\[
\Iket{B_n}_{T_{g_1}}^{(0)} \sim \zetabar^n\ket{+}\otimes\ket{+} +
i\eta\ket{-}\otimes\ket{-}\,, 
\]
up to normalisations.  The action of $g_2$ on these states is given by
the action of $g_2$ on the oscillators, and the action on the state we
have defined as our ground state, namely the state
$\ket{-}\otimes\ket{-}$. From the discussion of the $D(r,0)$-brane we
know that the action of $g_2$ therefore is given by
\[
g^k_2 (\zetabar^n\ket{+}\otimes\ket{+} + i\eta \ket{-}\otimes\ket{-}) =
\omega^k \zetabar^k 
(\zetabar^n\ket{+}\otimes\ket{+} + i\eta \ket{-}\otimes\ket{-}) \,,
\]
where $\omega$ is the discrete torsion phase by which we have to
modify the action of $g_2$ on the ground state in the $g_1$-twisted
sector. Thus if we define the RR sector ground states as
\begin{equation}
  \Iket{B_n}_{T_{g_1}}^{(0)} =
  \omega^n\zetabar^n\left(\zetabar^n\ket{+}\otimes\ket{+} + i\eta\ket{-}\otimes\ket{-}\right)\,,
\end{equation}
then $g_2\Iket{B_n}_{T_{g_1}}^{(0)} =
\Iket{B_{n+1}}_{T_{g_1}}^{(0)}$. 
By summing over the ground states $\sum_n
\Iket{B_n}_{T_{g_1}}^{(0)}$ we see that $i\eta\ket{-}\otimes\ket{-}$ 
survives, if the discrete torsion phase is given by $\omega=\zeta$, and
$\ket{+}\otimes\ket{+}$ survives if $\omega=\zetabar$. Thus we have
constructed a brane which couples to the twisted RR sector.

We now have to construct the open strings corresponding to these
branes. Since the orbifold invariant combination of branes is a
superposition of the three boundary states $\Iket{B_n}$, the open
string spectrum has sectors which correspond to strings that start and
end on the branch $\Iket{B_n}$ of the orbifold invariant combination,
and open strings which stretch between different branches $\Iket{B_n}$
and $\Iket{B_m}$.  The open string which corresponds to this boundary
state therefore has a $3\times3$ Chan-Paton matrix
\begin{equation}
  \label{eq:6}
  A:=
  \begin{pmatrix}
    a_{00} & a_{01} & a_{02} \\
    a_{10} & a_{11} & a_{12} \\
    a_{20} & a_{21} & a_{22} 
  \end{pmatrix}\,,
\end{equation}
where the diagonal elements label strings that start and end on the
same brane, and the off-diagonal elements label strings which stretch
between different branes. In this notation, $a_{ij}$ labels a string
which starts at brane $i$ and ends at brane $j$, and $a_{ji}$ labels a
string with the opposite orientation. The orbifold group acts on the
Chan-Paton matrix by conjugation
\[
A\mapsto\gamma(g_k)A\gamma(g_k)^{-1}\,.
\]

From the action of the $g_i$ on the boundary state, and from the
modular transformation of the cylinder amplitudes, we can read off the
action of $g_1$ and $g_2$ on the Chan-Paton factors. They are given by
\begin{align}
  \label{eq:45}
  \gamma(g_1) &= 
  \begin{pmatrix}
    1 &0&0 \\
    0 & \omega & 0 \\
    0 & 0 & \omega^2
  \end{pmatrix}\,,
  && \gamma(g_2) = 
  \begin{pmatrix}
    0 & 0 & 1 \\
    1 & 0 & 0 \\
    0 & 1 & 0
  \end{pmatrix}\,.
\end{align}
Thus $\gamma(g_1)$ multiplies strings stretching between
different branes by an appropriate phase, and $\gamma(g_2)$ permutes
strings on different branes. 
In principle we have to determine the action of $(-1)^F$ on the
Chan-Paton factors as well, but since the branes are not orthogonal or
parallel to each other this is hard to do \cite{CG}.

\subsection{Supersymmetry}
\label{sec:supersymmetry}

Although the gluing conditions we have considered might seem
unfamiliar, they actually describe conventional D3-branes. The most
general gluing conditions for free bosons and fermions can be written
as
\begin{equation}
  \label{eq:general-gluing}
  \left(a_r^i + M^{ij} \tilde{a}_{-r}^j\right) \Bket{B}=0 \,,
\end{equation}
where the matrix $M$ is some element of O($d$), $d$ being the
number of spacetime coordinates. If $M$ has eigenvalues different from
$\pm1$, this can be linked to electric fields on the brane.
Let us denote by $M_n$ the gluing matrix associated to $\Bket{B_n}$.
Then, $M_n$ is in terms of the real coordinates given by  
\begin{equation}
  \label{eq:gluing}
M_n=
  \begin{pmatrix}
    c_{2n} & s_{2n}  & 0 &0 & 0 & 0\\
    s_{2n} & -c_{2n} & 0 &0 & 0 & 0\\
    0 & 0 & 0 & 0 & c_{2n} & s_{2n}\\
    0 & 0 & 0 & 0 & s_{2n} &-c_{2n}\\
    0 & 0 & c_{2n} & s_{2n} & 0 & 0\\
    0 & 0 & s_{2n} &-c_{2n} & 0 & 0
  \end{pmatrix}\,,
\end{equation}
where we have defined
\begin{equation}
  c_m := \cos \left(\frac{m \pi}{3}\right) \qquad \text{and} \qquad
  s_m := \sin\left(\frac{m \pi}{3}\right) \,. 
\end{equation}
If we diagonalise $M_0$, we can verify that \eqref{eq:gluing} corresponds to a
$D3$-brane. In terms of the basis
\begin{align}
\label{eq:new-basis}
  b_r^1 &= a_r^1 \,, \nonumber \\
  b_r^2 &= a_r^2 \,,\nonumber\\
  b_r^3 &= \frac{1}{\sqrt{2}}\left(a_r^3+ a_r^5\right) \,,\\
  b_r^4 &= \frac{1}{\sqrt{2}}\left(a_r^4+ a_r^6 \right)\,,\nonumber\\
  b_r^5 &= \frac{1}{\sqrt{2}}\left(a_r^4 - a_r^6 \right)\,,\nonumber\\
  b_r^6 &= \frac{1}{\sqrt{2}}\left(-a_r^3 + a_r^5 \right) \,, \nonumber
\end{align}
$M_0$ takes the form $M_0=\text{diag}(1,-1,1,-1,1,-1)$. With respect to
this basis $M_n$ is given by
\begin{equation}
  \label{eq:rotated-gluing}
M_n =
   \begin{pmatrix}
    c_{4n}& s_{4n}& 0& 0& 0& 0\\
    s_{4n}& -c_{4n}& 0& 0& 0& 0\\
    0& 0&c_{2n}& -s_{2n}& 0& 0\\
    0& 0&-s_{2n}& -c_{2n}& 0& 0\\
    0& 0& 0& 0 &c_{2n}& -s_{2n}\\
    0& 0& 0& 0 &-s_{2n}& -c_{2n}
  \end{pmatrix}\,. \\
\end{equation}
All $M_n$'s have three eigenvalues $(+1)$ and three eigenvalues
$(-1)$. They therefore correspond to D3-branes without flux. 

In terms of the $b_r^i$ the action of $g_1$ is given by a rotation by
an angle of $\tfrac{2 \pi}{3}$ in the $b^3$-$b^5$ and the $b^4$-$b^6$
plane. It can be easily checked that indeed $g_1^{-k}M_ng_1^{k} = M_n$.
The action of $g_2$ in the $b^i$ basis is given by the matrix 
\begin{equation}  
g_2=
\begin{pmatrix}
 c_{2}& s_{2}& 0& 0& 0& 0\\
-s_{2}& c_{2}& 0& 0& 0& 0\\
0& 0& (c_{1})^2& -s_{1}c_{1}& s_{1}c_{1} & -(s_{1})^2\\
0& 0& s_{1}c_{1}& (c_{1})^2 & (s_{1})^2 & s_{1}c_{1}\\
0& 0& -s_{1}c_{1}& (s_{1})^2 & (c_{1})^2& -s_{1}c_{1}\\
0& 0 & -(s_{1})^2& -s_{1}c_{1}& s_{1}c_{1}&  (c_{1})^2
\end{pmatrix}\,.
\end{equation}
This matrix can be decomposed into commuting matrices $g_2^H$ and
$g_2^M$, 
\[g_2=g_2^Mg_2^H\,.\] 
$g_2^H$ leaves the world-volumes of the branes described by $M_n$ invariant
\begin{equation}
  (g_2^H)^{-m}M_n(g_2^H)^m = M_n\,,
\end{equation}
whereas $g_2^M$ rotates the world-volumes
\begin{equation}
    (g_2^M)^{-m}M_n(g_2^M)^m = M_{n+m}\,.
\end{equation}
Explicitly, $g_2^M$ and $g_2^H$ are given by
\begin{equation}
  g_2^H = 
  \begin{pmatrix}
    1 & 0 & 0 & 0 & 0 & 0\\
    0 & 1 & 0 & 0 & 0 & 0\\
    0 & 0 & c_{1}& 0 & s_{1} & 0 \\
    0 & 0 & 0& c_{1}& 0 & s_{1}  \\
    0 & 0 & - s_{1} & 0 & c_{1}& 0 \\
    0 & 0 & 0& - s_{1} & 0& c_{1} \\
  \end{pmatrix}
\end{equation}
and
\begin{equation}
  \label{eq:g2e}
  g_2^M = 
  \begin{pmatrix}
     c_{2}& s_{2}& 0& 0& 0& 0\\
     -s_{2}& c_{2}& 0& 0& 0& 0\\
     0 & 0 & c_{1}& -s_{1}& 0& 0\\
     0 & 0 & s_{1}& c_{1}& 0 & 0 \\
     0& 0& 0& 0 & c_{1}& -s_{1} \\
     0& 0& 0& 0 & s_{1}& c_{1} \\
  \end{pmatrix}\,.
\end{equation}
With respect to the $b^i$ basis each $M_n$ has one Neumann and one
Dirichlet direction in each of the planes $b^1$-$b^2$, $b^3$-$b^4$ and
$b^5$-$b^6$, respectively. $g_2^M$ acts as a rotation in each of these
three planes. 

After these preparations we are now in the position to apply the
supersymmetry analysis for branes at angles of
\cite{Berkooz:1996km,polchinski}. As is explained there, such
configurations of D3-branes preserve supersymmetry if the total angle
of rotations in $g_2^M$ is an integer multiple of $2\pi$, as is easily
checked. This shows that the above system of D3-branes is indeed
supersymmetric. 

\section{Conclusions}
\label{sec:conclusions}

We have constructed a supersymmetric stable BPS brane in an orbifold
with discrete torsion which couples to twisted and untwisted RR
potentials.  We called the brane constructed here a permutation
brane but, as explained in the previous section, in the
theory of free bosons and fermions permutation boundary states do not
form a new class of boundary states.

Even though we chose for the sake of concreteness to look at the
$g_1$-twisted sector, we could have constructed this kind of brane as
well for the $g_2$ or $g_1g_2$-twisted sectors, and the analysis would
have been identical. In each case the boundary state is a
supersymmetry preserving superposition of D3-branes.  We have
therefore managed to construct BPS D-branes which couple to the
twisted RR potentials.

Although we performed this construction explicitly for the case of a
$\mZ_3\times\mZ_3$ orbifold, it should be straightforward to generalise
it to other $\mZ_m\times\mZ_m$ orbifolds with odd $m$; the case with
even $m$ is different and has been solved in \cite{CG}.
For the sake of simplicity, we have only considered the uncompactified
theory, however, compactification on tori should be an easy
exercise. Another interesting test for mutual consistency would be to
calculate the amplitude between a conventional brane and the
brane constructed here.

\section*{Acknowledgements}

I am indebted to my supervisor Matthias Gaberdiel for his advice and
ideas, and a critical reading of the manuscript.  Furthermore I would
like to thank Ilka Brunner, Stefan Fredenhagen and Peter Kaste for
stimulating discussions. This research is supported by a scholarship
of the Marianne und Dr. Fritz Walter Fischer-Stiftung and a
Promotionsstipendium of the Deutscher Akademischer Austauschdienst
(DAAD) and the Swiss National Science Foundation.

\appendix

\section{Fermionic zero modes}
\label{sec:fermionic-zero-modes}

We use the following conventions for the fermionic zero
modes. $\psi^i_0$ and $\tpsi^i_0$ are defined to be raising operators,
and $\psi^{\ibar}_0$ and $\tpsi^{\ibar}_0$ are lowering operators. The
state annihilated by $\psi^{\ibar}_0$ we call $\ket{-}$, and the state
which arises by acting with the raising operator on $\ket{-}$ we call
$\ket{+}:= \psi^i_0\ket{-}$. We have to build tensor products of left
and right-movers and of various copies of the above algebra. Let us
start with the tensor product of left and right-movers. We denote 
the state $\ket{-}\otimes\ket{-}$ to be the state annihilated by both
$\psi^{\ibar}_0$ and $\tpsi^{\ibar}_0$. Furthermore, we define the
state
\[
\ket{+}\otimes\ket{+} := \psi^i_0\tpsi^i_0\ket{-}\otimes\ket{-}. 
\]
Due to the fact that fermions anti-commute, the order of the operators
is essential. We will order the raising operators such that the tilded
modes stand right of the untilded modes. 

Let us now consider the case where we have to pairs of raising left
moving creators and annihilators. Let us call them $\psi^1_0,\ 
\psi^2_0,\ \psi^{\bar{1}}_0,\ \psi^{\bar{2}}_0$. The ground state
shall be denoted by $\ket{--}$. Then we define the state $\ket{++}$ to
be given by 
\[
\ket{++}:=\psi^1_0\psi^2_0\ket{--} = - \psi^2_0\psi^1_0\ket{--}\,.
\]
Our conventions for the right-movers are identical. The case for more
than two oscillators on the left and right moving side is analogous.
With these conventions we obtain a definite sign rule which is used,  
wherever the RR ground states are considered.

\end{document}